\documentclass[conference]{IEEEtran}
\IEEEoverridecommandlockouts
\usepackage{amsmath,amssymb,amsfonts}
\usepackage{physics}
\usepackage{algorithmic}
\usepackage{graphicx}
\usepackage{textcomp}
\usepackage{hyperref}
\usepackage{enumitem}
\usepackage{fancyhdr}
\usepackage{ragged2e}
\usepackage{floatrow}
\usepackage{xcolor}
\usepackage{listings}
\usepackage{caption}
\usepackage[frozencache,cachedir=.]{minted}
\setminted[python]{ %
    linenos=true,             
    autogobble=true,          
    frame=lines,
    framesep=2mm,
    fontsize=\footnotesize
}
\makeatletter
\newenvironment{code}
 {\RecustomVerbatimEnvironment{Verbatim}{BVerbatim}{}%
  \def\FV@BProcessLine##1{%
    \hbox{%
      \hbox to\z@{\hss\theFancyVerbLine\kern\FV@NumberSep}%
      \FancyVerbFormatLine{##1}%
    }%
  }%
  \VerbatimEnvironment
  \setbox\z@=\hbox\bgroup
  \begin{minted}{python}}
 {\end{minted}\egroup
  \leavevmode\vbox{\hrule\kern2mm\box\z@\kern2mm\hrule}}
\makeatother

\usepackage{tikz}
\usetikzlibrary{quantikz}
\usepackage{adjustbox}
\def\BibTeX{{\rm B\kern-.05em{\sc i\kern-.025em b}\kern-.08em
    T\kern-.1667em\lower.7ex\hbox{E}\kern-.125emX}}
\usepackage[
backend=biber,
style=numeric,
url=false,
isbn=false,
eprint=false,
doi=false,
sorting=none
]{biblatex}
\newcommand{\doiorurl}{%
  \iffieldundef{doi}
    {\iffieldundef{url}
       {}
       {\strfield{url}}}
    {http://dx.doi.org/\strfield{doi}}%
}

\newcommand{\myhref}[1]{%
 \ifboolexpr{%
   test {\ifhyperref}
   and
   not test {\iftoggle{bbx:url}}
   and
   not test {\iftoggle{bbx:doi}}
  }
  {\href{\doiorurl}{#1}}
  {#1}
}

\DeclareFieldFormat{title}{\myhref{\mkbibemph{#1}}}
\DeclareFieldFormat
  [article,inbook,incollection,inproceedings,patent,thesis,unpublished]
  {title}{\myhref{\mkbibquote{#1\isdot}}}
\newcommand{\ms}{\,\text{ms}}

\author{

    \IEEEauthorblockN{
        Tobias Schmale\IEEEauthorrefmark{1}\textsuperscript{1}, 
        Bence Temesi\IEEEauthorrefmark{1}\textsuperscript{2}, 
        Niko Trittschanke\IEEEauthorrefmark{1},
        Nicolas Pulido-Mateo\IEEEauthorrefmark{2}\IEEEauthorrefmark{3},
        Ilya Elenskiy \IEEEauthorrefmark{5},\\
        Ludwig Krinner\IEEEauthorrefmark{2}\IEEEauthorrefmark{3},
        Timko Dubielzig\IEEEauthorrefmark{2},
        Christian Ospelkaus\IEEEauthorrefmark{2}\IEEEauthorrefmark{3},
        Hendrik Weimer\IEEEauthorrefmark{1}\IEEEauthorrefmark{4},
        Daniel Borcherding\IEEEauthorrefmark{1}\textsuperscript{3}
    }\\
    \IEEEauthorblockA{
        \IEEEauthorrefmark{1}\textit{Institut für Theoretische Physik, }
        \textit{Leibniz Universität Hannover, }
        Appelstraße 2, 30167 Hannover, Germany 
    }
    \IEEEauthorblockA{
        \IEEEauthorrefmark{2}\textit{Institut für Quantenoptik, }
        \textit{Leibniz Universität Hannover, }
        Welfengarten 1, 30167 Hannover, Germany
    }
    \IEEEauthorblockA{
        \IEEEauthorrefmark{3}\textit{Physikalisch-Technische Bundesanstalt, }
        Bundesallee 100, 38116 Braunschweig, Germany
    }
    \IEEEauthorblockA{
        \IEEEauthorrefmark{4}\textit{Institut für Theoretische Physik, }
        \textit{Technische Universität Berlin, }
        Hardenbergstraße 36, 10623 Berlin, Germany
    }
    \IEEEauthorblockA{
        \IEEEauthorrefmark{5}\textit{Institut für Elektrische Messtechnik und Grundlagen der Elektrotechnik, }
        \textit{Technische Universität Braunschweig, }\\
        Hans-Sommer Strasse 66, 38106 Braunschweig, Germany
    }
}

\begin{document}

\title{Real-time hybrid quantum-classical computations \\for trapped-ions with Python control-flow }

\maketitle

\begin{abstract}
In recent years, the number of hybrid algorithms that combine quantum and classical computations has been continuously increasing. These two approaches to computing can mutually enhance each others' performances thus bringing the promise of more advanced algorithms that can outmatch their pure counterparts. In order to accommodate this new class of codes, a proper environment has to be created, which enables the interplay between the quantum and classical hardware.

For many of these hybrid processes the coherence time of the quantum computer arises as a natural time constraint, making it crucial to minimize the classical overhead. For ion-trap quantum computers however, this is a much less limiting factor than with superconducting technologies, since the relevant timescale is on the order of seconds instead of microseconds. In fact, this long coherence time enables us to develop a scheme for real-time control of quantum computations in an interpreted programming language like Python.
In particular, compilation of all instructions in advance is not necessary, unlike with superconducting qubits.
This keeps the implementation of hybrid algorithms simple and also lets users benefit from the rich environment of existing Python libraries. 

In order to show that this approach of interpreted quantum-classsical computations (IQCC) is feasible, we bring real-world examples and evaluate them in realistic benchmarks.
\end{abstract}

\begin{IEEEkeywords}
Quantum Computing, Hybrid Quantum-Classical Computation, Trapped-Ions
\end{IEEEkeywords}

\footnotetext[1]{tobias.schmale@itp.uni-hannover.de}
\footnotetext[2]{bence.temesi@itp.uni-hannover.de}
\footnotetext[3]{daniel.borcherding@itp.uni-hannover.de}

\section{Introduction} 

Quantum computation as a field has been steadily growing in recent years as it holds the potential of solving problems exponentially faster than classical computers. Certain tasks have been theoretically proven \cite{shor1994algorithms, grover1996fast, bravyi2018quantum, rebentrost2014quantum}, or even demonstrated on quantum hardwares \cite{arute2019quantum, riste2017demonstration} to be faster than classical counterparts,
however achieving quantum advantage in practical problems has yet to be realized.

While further innovations on the hardware level are needed to realize quantum advantage, it is also possible to improve quantum computations by enhancing them with classical computations. This is particularly important in the current noisy intermediate-scale quantum (NISQ) era, when all the existing quantum hardwares are limited both by their noisy gates and by the number of qubits they have. Various algorithms demonstrate the advantage of such a hybrid quantum-classical computational model: The iterative phase estimation (IPE) requires far less qubits than the quantum phase estimation algorithm (QPE) by reusing one ancilla qubit together with intermediate classical computations \cite{griffiths1996semiclassical}. Quantum error correction requires classical look-up tables in order to determine correction protocols based on syndrome measurements.
Other algorithms like variational quantum algorithms (VQA) solve optimization problems classically to reduce the amount of entangling gates in a quantum circuit \cite{cerezo2021variational}. For an overview of hybrid quantum-classical algorithms, see \cite{lubinski2022advancing}.

However, on many quantum hardware approaches, implementing such hybrid algorithms is technically very challenging, since the coherence time of the physical qubits places a natural limit on the communication time between a quantum and a classical device. For systems such as superconducting or spin qubits, this time-limit is on the order of microseconds \cite{PhysRevLett.129.030501} -- meaning that any classical logic has to be compiled and executed on silicon with nanosecond latency access to the quantum device \cite{corcolesExploitingDynamicQuantum2021}. This places a burden on the quantum software developer, who has to write algorithms in a manner compatible with these restrictions. 

In many cases, entirely new quantum programming languages have been put forward to solve this problem. IBM recently added a simple classical control-flow to its Python-based software development kit \textsc{Qiskit} to support simple classical computations during a quantum computation based on OpenQASM 3.0 \cite{cross2022openqasm}.  Compiled languages such as Q\# with the intermediate representation QIR \cite{microsoft2021} as well as the language QCOR based on C/C++ \cite{mintz2020} have been developed 
as more complex alternatives.

Trapped-ions, on the other hand, offer significantly longer coherence times than the previously mentioned approaches, even reaching the time-scale of minutes \cite{wangSingleIonQubit2021, pinoDemonstrationTrappedionQuantum2021}. This makes some of the so far proposed solutions for quantum-classical programming disproportionate, as they solve a problem that does not exist for trapped-ion quantum computers. In this paper, we exploit these long coherence times to propose a new user interface for \emph{interpreted} quantum-classical computations (IQCC) based on standard Python control-flow.
This allows developers of quantum software to benefit from the simplicity of the Python language and the abundance of libraries and software-frameworks available for it. Furthermore, although many specific-purpose quantum languages exist, Python remains one of the most popular languages for developing static, pure quantum algorithms. Therefore, our proposal can make use of all of this existing work and extend it to dynamic real-time quantum-classical software.

Our framework for Python-based control-flow enables the implementation of all types of quantum-classical algorithms prominent today, including fast quantum error correction all the way to variational quantum algorithms. 
We demonstrate that our software setup is feasible by evaluating and benchmarking multiple real-world examples and comparing to execution speed expected for upcoming trapped-ion quantum computers, e.g. the QVLS-Q1 chip \cite{schmale2022backend}.

\section{Concept}
\label{sec:concept}
\subsection{User API}
\label{sec:user_api}
To combine quantum and classical computations we use an existing Python package to generate quantum circuit objects and combine it with the standard Python control flow. That way users have the ability to generate quantum circuit objects and at the same time have the full flexibility of Python together with external packages to enhance their quantum algorithms with classical computations. In the following we use the software development kit \textsc{Qiskit} \cite{aleksandrowicz2019qiskit}, as an example to generate quantum circuit objects. However, any other Python package for quantum circuits with the ability to export circuits in the language OpenQASM 2.0 would be suitable.

In our framework a quantum-classical program is a Python script that triggers the execution of a quantum circuit (\textsc{Qiskit} object quantum\_circuit) on a quantum device in real-time by calling the function \texttt{qpu.run(quantum\_circuit)}. The result of a hybrid algorithm can be uploaded to a database by calling the function \texttt{upload()} on the result object. Fig.~\ref{fig:hybrid_hello_world} shows the main structure of a quantum-classical program with two QPU calls (line 5 and 9) and placeholders for classical computations that also generate the quantum circuit objects that are being executed. In line 11 the result object is being uploaded.

\begin{figure}[!ht]
\begin{code}
from qvls import qpu, upload

# Insert classical computations
    
result = qpu.run(quantum_circuit_1)
    
# Insert more classical computations
    
result = qpu.run(quantum_circuit_2)

upload(result)

\end{code}
\caption{Example of a hybrid algorithm that triggers the execution of a quantum circuit on a quantum device in real-time (line 5 and 9). The placeholders indicate where classical computations that eventually generate the quantum circuit objects can be added. In the end the result object is uploaded to a database.}
\label{fig:hybrid_hello_world}
\end{figure}

\subsection{Examples}
Using the terminology from \cite{ibmhybridjobs2022}, we distinguish between real-time hybrid computations that apply classical computations within the coherence time of the qubits, and near-time hybrid computations that reset the quantum computer before each run. Here, we show examples that demonstrate that our user interface is applicable to both types of hybrid computations.

A simple example of a real-time quantum-classical computation is the three qubit bit flip code \cite{nielsen2002quantum}, shown in Fig.~\ref{fig:bit_flip_code}. In this example the state $\ket{000}$ is prepared and a Pauli $X$ error (line 6) is applied to the first qubit. Afterwards, two ancilla qubits are taken to perform the parity (or ``stabilizer") measurements $Z_0Z_1$ and $Z_1Z_2$ (lines 8-14). The results of the stabilizer measurements are then used in a classical condition to correct the error (lines 16-23) such that the final state should be $\ket{000}$ and the hybrid program should return the corresponding dictionary \{``000'' : 1\} (line 29). 

\begin{center}
\begin{figure}[!ht]
\begin{code}
from qvls import qpu, upload
from qiskit import QuantumCircuit

qc = QuantumCircuit(5, 3)
    
qc.x(0)
    
qc.cx(0,3)
qc.cx(1,3)
qc.measure(3,0)

qc.cx(1,4)
qc.cx(2,4)
qc.measure(4,1)

result = qpu.run(qc)

if '001' in result:
    qc.x(0)
elif '010' in result:
    qc.x(2)
elif '011' in result:
    qc.x(1)

qc.measure_all()

result = qpu.run(qc)

upload(result)
\end{code}
\caption{Three qubit bit flip code implemented using our Python control-flow.}
\label{fig:bit_flip_code}
\end{figure}
\end{center}

Our next example showcases how the user can run near-time hybrid jobs on the hardware, such as VQAs. These algorithms can generally be summarized in the following high-level steps: 
\begin{enumerate}
    \item Creating a parametrized ansatz circuit for the problem at hand. \vspace{0.1cm}
    \item Choosing a cost function, and measuring it with the quantum computer for each evaluation. \vspace{0.1cm}
    \item Updating the parameters with some method (gradient-based or not), using a classical computer, such that the measured cost function is optimized.
\end{enumerate} \vspace{-0.3cm}
\begin{center}
\begin{figure}
\begin{code}
from qvls import qpu, upload
from scipy.optimize import minimize

# Create ansatz circuit
ansatz_circuit = create_circuit()

# Compile symbolically
circuit = qpu.compile(ansatz_circuit)

def cost_func(params):
  # Measure and compute cost function
  result = backend.run(circuit, params)
  return compute_cost(result)

# Use classical optimizer to find solution
results = minimize(cost_func, initial_params)
upload(results)
\end{code}
\caption{Example of a near-time hybrid job, here a VQA, that recurrently measures the cost function with the quantum computer, and optimizes the parameters classically.} \label{fig:vqe}
\end{figure}
\end{center}
The code snippet in Fig.~\ref{fig:vqe} gives a general outline of how we execute such VQAs. In addition to the steps previously mentioned, we can notice the followings: we use \texttt{qpu.compile()} to transpile and compile the parametrized circuit once, so that during the optimization the updated parameter values only have to be assigned, thus avoiding redundant compilations. Also, in line 16, we see how one can straightforwardly use any kind of optimizers taken from the Python package of their choice, such as \textsc{Scipy} \cite{2020SciPy-NMeth}.

To both real- and near-time hybrid jobs we will give specific, physically relevant examples and benchmark them in Sec.~\ref{sec:evaluation}.
\subsection{Software- and server architecture}
\label{sec:serverarchitecture}

In this section we discuss the server architecture that is needed to realize the user API introduced in section~\ref{sec:user_api} for real-time and near-time quantum-classical computations. In the real-time case the main requirement of a hybrid algorithm with total runtime $t_{\text{total}}$ is that it finishes within the coherence time $t_{\text{coherence}}$ of the ions in the quantum hardware, i.e. $t_{\text{total}} < t_{\text{coherence}}$ is required. For any hybrid algorithm we can contribute to shorter total runtimes by reducing the overhead of our setup. In particular, it is desirable to 
ensure the total runtime of the hybrid algorithm is dominated by the quantum contribution $t_\text{quantum}$, and not by classical computations and other classical contributions $t_{\text{classical}}$. This ensures the more valuable quantum resource is used efficiently. 

We therefore quantify the efficiency of our setup for a particular algorithm using the classical overhead, defined as $t_\text{classical}/t_\text{total}$, where $t_\text{total} = t_\text{classical} + t_\text{quantum}$.

Another requirement appears when the quantum computer is available to unknown users: in this case one needs to ensure that the untrusted Python code of a quantum-classical job can be executed securely. Hence, we end up with the following goals for the server architecture:
\vspace{0.2cm}

\begin{enumerate}
    \item Performance: Minimize the classical overhead $t_{\text{classical}} / t_{\text{total}}$ for hybrid algorithms.\vspace{0.1cm}
    \item Security: Secure execution of untrusted Python code.
\end{enumerate}
\vspace{0.2cm}

To meet these requirements, we chose a server architecture consisting of a QPU part (FPGA and trapped-ions), a Mediator computer (``Experimental control system") with an interface to the QPU and another computer called CPU that runs the untrusted Python code inside a sandboxed environment (e.g. a docker container). In this architecture the Mediator mediates between the QPU and the CPU: On the one hand it can receive assembly instructions (in our case TIASM instructions \cite{schmale2022backend}) from the CPU, converts it into FPGA instructions and forwards them to the FPGA part of the QPU. On the other hand it can receive measurement results from the QPU and forwards them to the CPU.
All of these participants are located on the same local network.
In principle the software for the Mediator and the CPU could run on the same physical device, however we chose to further protect the QPU from the execution of untrusted Python code. 

\begin{figure}
    \centering
    \includegraphics[width=0.95\textwidth]{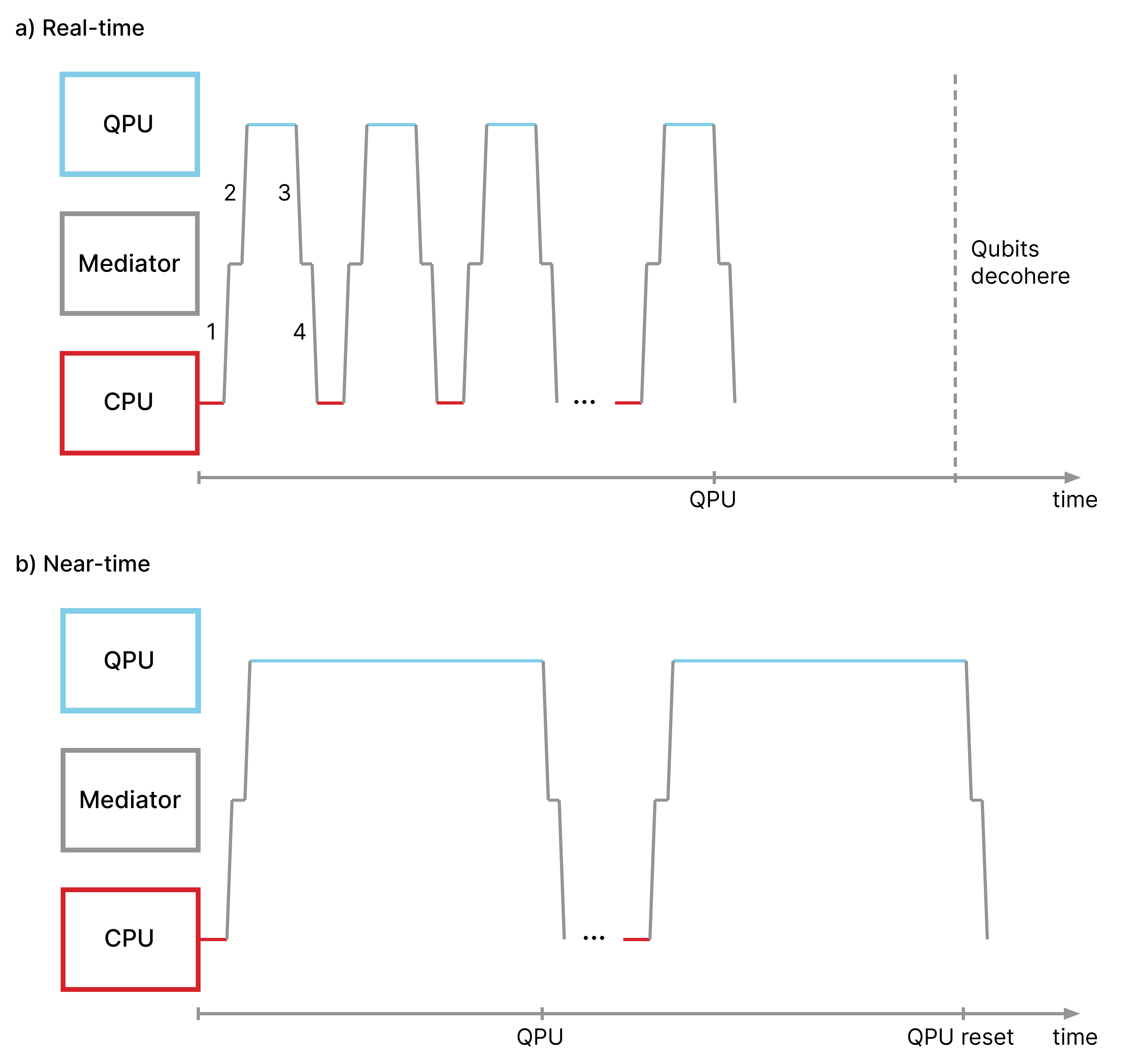}
    \caption{Comparison of the time spent at the CPU, the Mediator and the QPU (FPGA and trapped-ions) (horizontal lines) during a hybrid algorithm. The numbers label different communication steps: 1) Sending TIASM from CPU to Mediator, 2) sending pulse information from Mediator to QPU, 3) sending result information from QPU to Mediator and 4) sending measurement results from Mediator to CPU. Notice that in the real-time case the hybrid algorithm finishes before the qubits decohere while in the near-time case the quantum computer is reset after each cycle.}
    \label{fig:realtime_neartime}
\end{figure}

This is the simplest way of dividing up the server architecture for hybrid computations to meet the security constraints. Below we show that it turns out to also be sufficient for achieving the performance goals. 
Fig.~\ref{fig:realtime_neartime} shows the communication and the expected timings between all participants in the network for near-time and real-time algorithms.

\section{Evaluation}
\label{sec:evaluation}
\begin{figure*}
    \centering
    \includegraphics[width=\textwidth]{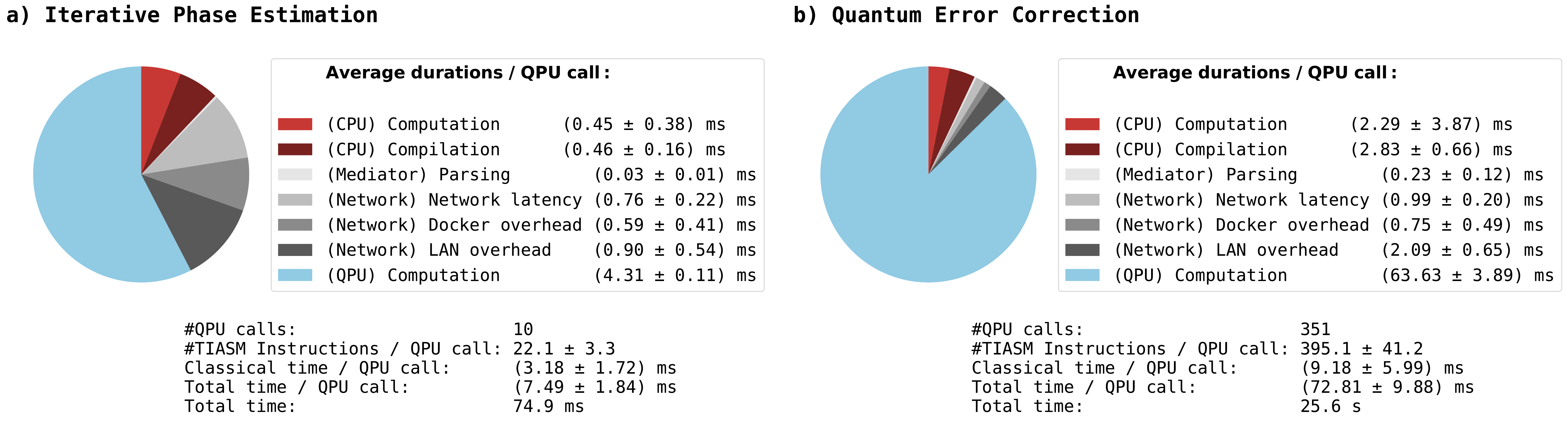}
    \caption{The contributions to the runtime are shown for the QEC and IPE example. The error bars indicate the standard deviation across the entire job, in order to show that the QPU calls made by real-world algorithms are not all the same across one algorithm. Detailed explanations of the individual contributions are given in Sec.~\ref{sec:benchmarking_strategy}}
    \label{fig:pie_charts}
\end{figure*}
In this section, we aim to demonstrate the feasibility of  real-time and near-time IQCCs for trapped-ion quantum computers, which means showing that all criteria listed in the section~\ref{sec:serverarchitecture} are met. Of most interest here is showing that the execution speed of hybrid algorithms is limited by the speed of the QPU and not by any classical overheads. As a specific example for a trapped-ion quantum computer we chose the upcoming QVLS-Q1 chip and its compiler within this benchmark \cite{schmale2022backend}.

\subsection{Examples used for benchmarking}
To do so, we run three hybrid quantum algorithms that are of real-world interest:

As a first example we consider Iterative Phase Estimation (IPE). 
This algorithm can estimate the eigenvalue of a unitary operation for a given eigenstate, and
is interesting as it appears as a sub-routine in other, larger algorithms, such as Shor's algorithm \cite{shor1994algorithms}. This is a relevant test case for our setup, as the circuits executed in each QPU call are very short, only consisting of $\sim\ 10$ native gates (in our example) that need to be executed before another classical decision on the CPU has to be made. Therefore, it is a worst case scenario for our setup, as short QPU times inherently imply a greater ratio of network time to QPU time, and can hence lead to a larger classical overhead. See Appendix \ref{appendix:ipe} for more details on IPE and \cite{corcolesExploitingDynamicQuantum2021} for a dynamic implementation using superconducting qubits.

Next, we consider Quantum Error Correction (QEC). QEC potentially holds the key to achieving quantum advantage. Most implementations, such as the Steane code which we use here, rely on parity measurements to detect error syndromes during a quantum computation, and use classical lookup tables to determine the quantum operations necessary to correct detected errors. Due to the embedding of logical qubits into many physical qubits, the workload per QPU call is significantly higher, in our case on the order of 100-200 quantum gates between two classical decisions. See Appendix \ref{appendix:steane} for more details and \cite{andersonRealizationRealTimeFaultTolerant2021a} for the first dynamic, experimental realization.
    
Lastly, to benchmark how VQAs run on our setup, we prepare the ground state of the 1D transverse Ising model for 7 qubits by using the Variational Quantum Eigensolver (VQE) \cite{peruzzo2014variational}. This hybrid algorithm can prove valuable for quantum simulations on near-term quantum computers, because it requires less amount of noisy quantum resources. VQEs find the ground state of a Hamiltonian by varying the parameters in the ansatz circuit, such that the energy (cost function) of the system is minimized. While the parameters are updated classically, evaluation of the energy happens by measuring the expectation value of the Hamiltonian with the quantum hardware. 
More details on the VQE implementation chosen here can be found in the Appendix \ref{appendix:vqe}.

All of these examples are made available for download under \cite{hybridalgorithms2023}, as minimal working examples in a local environment. Note that the details of the user API may be subject to change.

\begin{table}[]
    \centering
    \begin{tabular}{l|c}
         Operation & Assumed duration  \\\hline
         Single Qubit gate & 0.1\ms\ \cite{zarantanelloEntangling2019}  \\
         Two Qubit gate    & 1\ms\  \cite{zarantanelloEntangling2019}  \\
          Ion cooling       & 1\ms\  \cite{pinoDemonstrationTrappedionQuantum2021}\\
         Shutteling        & 0.1\ms\  \cite{burtonShuttelingTimes2022}  \\
         Measurement       & 0.1\ms\  \cite{pinoDemonstrationTrappedionQuantum2021}
    \end{tabular}
    \caption{\normalfont Durations assumed for various quantum operations. An ion cooling operation is assumed to take place before every two-qubit gate, leading to an effective two-qubit-gate-time of 2\,\ms. However, all these numbers should not be treated as exact values but just as orders of magnitude, since the exact numbers will depend on the concrete hardware implementation and calibration.}
    \label{tab:operations_times}
\end{table}

\subsection{Benchmarking strategy}\label{sec:benchmarking_strategy}
For benchmarking, we take into account the following contributions to the runtime of the algorithm:
\subsubsection{QPU execution time}
While we do not yet have access to a working QPU, we can estimate the runtimes of quantum circuits executed on the QPU by combining the compiled circuit instructions for the upcoming QVLS-Q1 chip \cite{schmale2022backend} with the durations for gates, measurements and ion movements from previous works. Table \ref{tab:operations_times} lists the assumed durations for various instructions of a trapped-ion chip. These durations are taken as given and optimizing them is outside of the control of our software stack. Using this, we compute the average QPU time for each job (``QPU - Computation" in Fig.~\ref{fig:pie_charts}). 

\subsubsection{Classical computation time} This is split up into the following components: 
Compilation, i.e. mapping the native gates onto the trapped-ion chip, performing the ion-routing and generating the assembly instructions for the QPU \cite{schmale2022backend}. And secondly, any remaining time spent on the classical computer ``CPU - Computation" in Fig.~\ref{fig:pie_charts}). This contains executing the user Python code in between two QPU calls, i.e. the classical part of the hybrid quantum algorithm.

\subsubsection{Experimental control-system overhead} This only includes the time required to parse the assembly instructions, as this is the only metric we currently have access to. In particular, any overhead in signal-generation are not included. However, we anticipate these overheads to be negligible compared to the other contributions.

\subsubsection{Network contributions} As described in the Server Architecture section~\ref{sec:serverarchitecture}, we isolate the user-submitted Python code from the experimental control system. This induces several contributions to network latency which we break down as follows:
\begin{itemize}
    \item Network latency: This is the pure latency induced by transmitting the QPU instructions over a TCP connection, when both communicating parties are running on the same host (the sum of 1-4 in Fig.~\ref{fig:realtime_neartime}).
   \item LAN overhead: This is the difference in total network latency that is added when moving one of the communicating parties to a separate host on the same network. We acknowledge, that this number strongly depends on the network setup. We tried to best replicate the anticipated future experimental setup, by simply connecting two machines via one switch on a 1 Gigabit/s ethernet network.
    \item Sandboxing overhead: This is the difference in total network latency that is added when isolating the user-submitted Python code in a docker container.  
\end{itemize}

\subsection{Benchmarking results}

The contributions of all of these components are shown in Fig.~\ref{fig:pie_charts} for the IPE and the QEC example. In both examples, the QPU is the largest contribution to the total runtime. 

For IPE, we achieve $t_\text{classical}/t_\text{total} \approx \frac{3.18\ms}{7.49\ms} \approx 42.5\,\%$, which reduces to $\approx 34.6\,\%$ if the LAN contribution can be neglected, as would be the case if a single machine were used. 
Although the classical overhead here is significant, it does not hinder the implementation's feasibility. The IPE algorithm is still short enough ($74.9\ms$ in our case) to be well below the coherence time of several seconds \cite{hahnTwoqubitMicrowaveQuantum2019}. This makes the absolute duration for which QPU time is ``wasted'' by classical overheads very short and thus the IQCC implementation practical.

For QEC, $t_\text{classical}/t_\text{total} \approx \frac{9.18\ms}{72.81\ms} \approx 12.7\,\%$, or $\approx 10.1\,\%$ in the single-host scenario.
Therefore, the total runtime of the algorithm is limited by the QPU and even if the classical overhead were eliminated entirely, the runtime could only be sped up by $10.1\,\%$. 

For VQE, it is not meaningful to plot a pie chart like for the previous examples, because the QPU time dominates the chart alone. The reason is that during the optimization the energy of the system is found by sampling over multiple measurements, i.e. the quantum circuit has to be run tens to thousands of times (number of shots) for a QPU call. In the previous real-time algorithms, a QPU call meant that the circuits were run just once, which made the quantum and classical time scales comparable. Here however, the QPU time is orders of magnitude larger than the classical times, $t_\text{classical} / t_\text{total} < 1\%$. Quantitatively: for measured classical overheads of $10-100\ms$, including transpilation time and LAN overhead, the cumulative QPU times are in the range of seconds, and even 10's of seconds. Note, that quantum error mitigation techniques could be utilized to improve the accuracy of results. In such a scenario, the classical overhead might become more significant than in our benchmarked example.
\begin{figure}
    \centering
    \includegraphics[width=\textwidth]{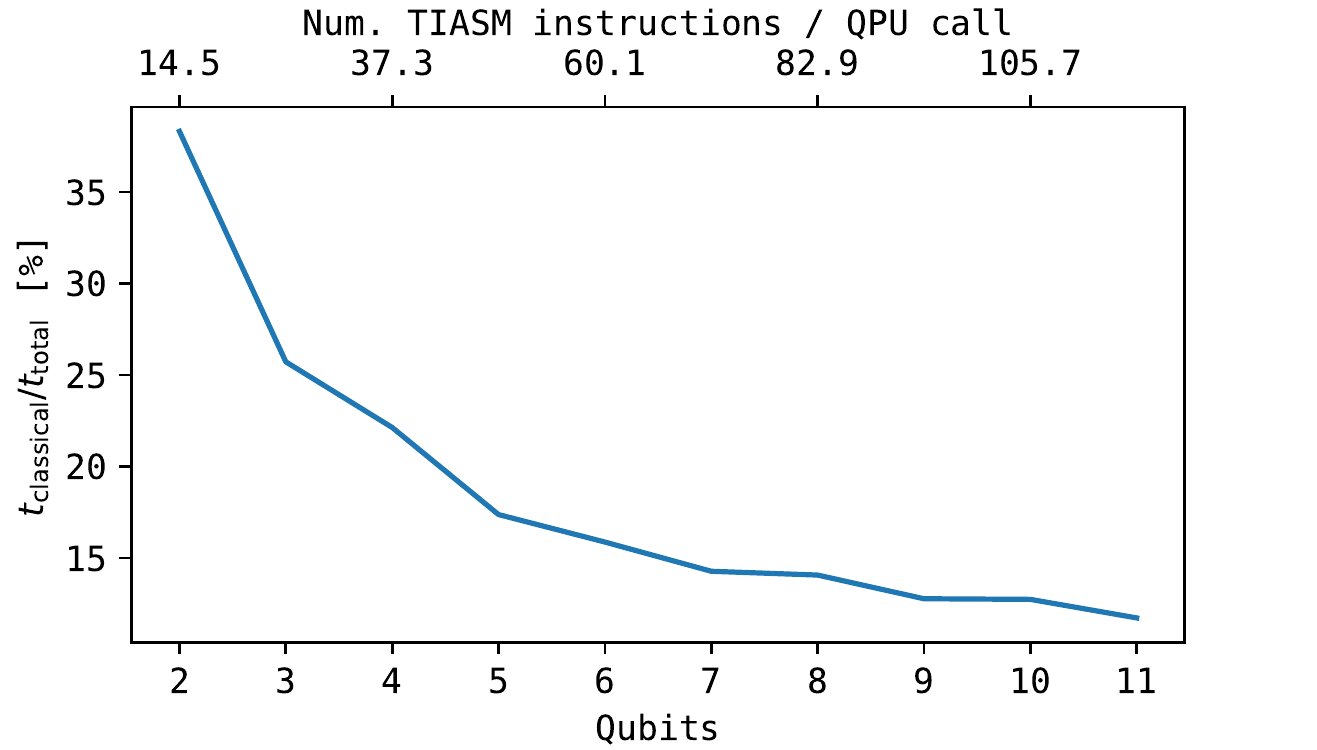}
    \caption{The classical overhead is shown as a function of qubit number in the IPE example. Here, more qubits implies more hardware instructions per QPU call, hence all contributions to the timing diagram take longer. However, the ratio of classical overhead to total time still decreases, i.e. IQCCs become more feasible the more complex the algorithm.}
    \label{fig:ratio_plot}
\end{figure}
With more complex QPU operations, the classical overhead scales better than the time spent in the quantum computer. We demonstrate this by benchmarking the IPE example with varied number of qubits that the relevant unitary acts on. This increases the durations of all classical contributions but also the duration of the QPU activity. The ratio of these can be seen in Figure~\ref{fig:ratio_plot}, confirming the expected trend and showing the setup is scalable in the number of qubits.

Hence, based on our benchmarks, we claim that all these types of algorithms are feasible when implemented using Python control-flow on a trapped-ion quantum computer.

\section{Conclusion and Outlook}
We have presented a user API and server architecture for implementing real-time and near-time hybrid quantum algorithms, that makes use of trapped-ions long coherence time to enable an interpreted, Python-based control flow. Through relevant hybrid examples, we have shown that the induced classical overhead is minor: for variational quantum algorithms it takes up less  than $1\,\%$ , and for quantum error correction on the order of $10\,\%$. Furthermore, tasks that require an even tighter interplay between quantum and classical hardware, such as iterative phase estimation, are also shown to be feasible.
Finally, in our most demanding example the classical overhead decreases when increasing the amount of qubits used by the hybrid algorithm. Thus, we have demonstrated that the interpreted hybrid workflow stays a viable choice even as the quantum volume of ion-trap hardware will increase.


In future works, we intend to investigate how ideas from OpenQASM3 \cite{cross2022openqasm} and IBMs dynamic circuits \cite{corcolesExploitingDynamicQuantum2021} can help to further decrease the remaining overhead. We imagine the user in the future to be able to choose between a compiled and an interpreted control-flow, and to be able to fluently switch between the two in the same quantum algorithm.

\section*{Acknowledgment}
This work was funded by the Quantum Valley Lower Saxony Q1 project (QVLS-Q1) through the Volkswagen foundation and the ministry for science and culture of Lower Saxony and by Germany’s Excellence Strategy – EXC-2123 QuantumFrontiers – 390837967.

\printbibliography

\appendix
\section{Appendix}

\subsection{Effect of transpilation}
In all the examples that were benchmarked here, it was assumed that the circuits were present in native gates. This is not an issue, as transpiling circuits is always possible before execution and modern software frameworks like \textsc{Qiskit} or \textsc{tket} \cite{pytket} even support this symbolically.

However, doing this \emph{pre-}transpilation requires one extra step, which we make easy by providing a decorator that handles this, e.g.:\\
\begin{center}
\begin{code}
@pretranspile
def circuit_part(alpha, beta)
    qc = QuantumCircuit(2)
    qc.rzz(alpha, 0, 1)
    qc.rx(beta, 0)
    
circuit = circuit_part(alpha=1, beta=2)
qpu.run(circuit)
\end{code}
\end{center}
Here, transpilation of the circuit defined in the function circuit\_part is transpiled in advance and subsequent calls to it very efficiently return the circuit in native gates, without any retranspilation.

If one does not use this, transpilation would have to occur within the coherent loop of the hybrid algorithm, which can take up to 10's of milliseconds, even for fast, C++ based packages such as \textsc{tket}. For the IPE example this would introduce a further overhead of $\sim3\ms$ and for the QEC example $\sim15\ms$.

\subsection{IPE algorithm}
\label{appendix:ipe}

In the IPE algorithm the phase $\phi$ of a unitary $U$ and an eigenstate $\ket{\psi}$, according to $U\ket{\psi} = \text{e}^{\text{i} \phi}\ket{\psi}$, is estimated by iteratively computing different parts of the binary expansion of $\phi$ \cite{griffiths1996semiclassical}. For the example in Section~\ref{sec:evaluation} we estimated the phase $\phi$ of the phase gate

\begin{equation}
P(\phi)=\begin{pmatrix}
    1 & 0 \\
    0 & \text{e}^{\text{i} \phi} 
\end{pmatrix}
\end{equation}
for the eigenstate $\ket{1}$. The main contribution of gates in the IPE algorithm comes from the consecutive execution of the controlled $U$ gate

\begin{equation}
    CU = I \otimes \ket{0}\bra{0} + U \otimes \ket{1} \bra{1}.
\end{equation}

In the case of the controlled phase gate $CP(\phi)$ the following identity can be used to reduce the amount of gates per QPU call:

\begin{equation}
    CP(\phi)^n = CP(n \phi).
\end{equation}

That way we ended up with $\sim10$ native gates for each QPU call. See \cite{hybridalgorithms2023} for the complete implementation. 


\subsection{Steane code}
\label{appendix:steane}
The Steane code is a quantum error correction code that requires seven physical qubits for one logical qubit and is capable of correcting arbitrary single-qubit errors. For the example in Section~\ref{sec:evaluation} we used the encoding circuit of \cite{Paetznick2012} and the $T$ gate implementation of \cite{Chamberland2019faulttolerantmagic} with the small adjustment of $T$ being defined as:

\begin{equation}
T=\begin{pmatrix}
    1 & 0 \\
    0 & \text{e}^{\text{i} \pi / 4} 
\end{pmatrix}.
\end{equation}

To demonstrate how frequently classical computations are needed we implemented a hybrid program that executes the following logical circuit
\begin{center}
\vspace{0.3cm}
\begin{adjustbox}{width=0.22\textwidth}
    \begin{quantikz}
        \lstick{$\ket{0}$}& \gate{RX(0.4)} & \ctrl{1} & \qw \\
        \lstick{$\ket{0}$}& \qw & \targ{} & \qw
    \end{quantikz}
\end{adjustbox}
\vspace{0.3cm}
\end{center}
on two logical qubits. Since the Steane code is only capable of executing Clifford + $T$ gates we used the Solovay-Kitaev algorithm with a recursion degree of two to transpile the example circuit to 137 logical Clifford + $T$ gates. On top of that, we measured all six stabilizers on both logical qubits after each execution of a logical gate, as is required in a fault-tolerant implementation of the Steane code \cite{nielsen2002quantum}. In total we ended up with approximately 100-200 native gates per QPU call to demonstrate the classical overhead of the Steane code. See \cite{hybridalgorithms2023} for all details of the implementation.

\subsection{VQE algorithm}
\label{appendix:vqe}

Here, we give more details on the VQE we used, and also elaborate on how different elements of a VQA can affect benchmarking results. With VQE we prepared the ground state of the following 1D transverse Ising Hamiltonian:
\begin{equation} \label{eq:ising_ham}
    H = - \sum_{\langle i, j\rangle}\sigma_i^z\sigma_j^z - h\sum_i\sigma_i^x,
\end{equation}
where $\{\sigma^\mu\}$ are the Pauli matrices, $h$ the coupling strength, and we are considering only nearest neighbour interactions $\langle\cdot , \cdot\rangle$. As the Hamiltonian has only two sets of non-commuting terms, we need to measure the circuit just twice for an energy evaluation. But many times we have to deal with more complicated Hamiltonians or with cost functions that need more measurements, which would increase the QPU times even more than what we have seen in our results.

Choosing a classical optimizer also has a direct impact on the number of QPU calls. In simulations we used the gradient-based Simultaneous Perturbation Stochastic Approximation (SPSA) optimizer \cite{spall1998overview}. One of its specialty is that it can approximate the gradient with just two evaluation of the cost function, independent of the number of parameters in the circuit. Thus, it represents the least amount of QPU-using option compared to other optimizers.

Finally, it depends on the ansatz circuit, how or if the VQA can converge to the solution. In our case, the Hamiltonian variational ansatz (HVA) \cite{wecker2015progress} prepares the variational state by parametrizing the terms in the Hamiltonian (1), i.e. a variational layer consists of $RZZ(\alpha)$ and $RX(\beta)$ gates, with $\alpha, \beta \in [0, 2\pi)$. For three qubits a HVA layer looks like the following:
\begin{center}
\vspace{0.3cm}
\begin{adjustbox}{width=0.45\textwidth}
    \begin{quantikz}
        \lstick{$q_0$}& \gate[2]{RZZ(\alpha_{01})} & \qw & \qw & \gate[3]{RZZ(\alpha_{02})} & \gate{RX(\beta)} & \qw\\
        \lstick{$q_1$}& & \gate[2]{RZZ(\alpha_{12})} & \qw & \qw & \gate{RX(\beta)} & \qw\\
        \lstick{$q_2$}& \qw & & \qw & & \gate{RX(\beta)} & \qw
    \end{quantikz}
\end{adjustbox}
\vspace{0.3cm}
\end{center}
In HVA, we have different parameters in a layer only for gates that do not commute with each other, e.g. in the circuit the $RZZ$ gates. Ideally, we want to keep the number of parameters low to avoid excessive cost function evaluations, but large enough to have the solution contained in the parameter space. However, dependent on the physics in the problem, the entanglement, or the system size, the number of variational layers, i.e. the number of parameters can grow significantly. All of which can suppress the classical overhead even more compared to the QPU times.


\end{document}